\documentclass{elsart}

\usepackage{epsfig}

\usepackage{amssymb}

\begin{document}

\begin{frontmatter}


\title{Testing the effective scale of Quantum Gravity with the next generation of Gamma Ray Telescopes}

\author{O.Blanch},
\ead{blanch@ifae.es}
\author{J.Lopez},
\ead{jlopez@ifae.es}
\author{M.Martinez}
\ead{martinez@ifae.es}

\address{IFAE, Barcelona, Spain}

\begin{abstract}
The actual potential of the next generation of Gamma Ray 
Telescopes in improving the existing tests of an effective Quantum
Gravity scale from the study of the propagation delay for gamma 
rays of different energies coming from a distant astrophysical 
source is discussed. 
It is shown that the existence of a cosmological Gamma Ray Horizon,
will impose very demanding conditions on the observations of the 
telescopes to try to test a Quantum Gravity scale close to the Planck 
mass.
\end{abstract}

\end{frontmatter}

\section{Introduction}

Imaging Cherenkov Telescopes have proven to be the most successful tool
developed so far to explore the cosmic gamma rays of energies above
few hundred GeV. A pioneering generation of installations has been able
to detect a handful of sources and start a whole program of very
exciting physics studies. Now a second generation of more elaborated 
telescopes is under construction and will provide soon with new 
observations. One of the main characteristics of these new telescopes
is the ability to detect lower energy gamma rays, which could fill the
observational gap between 10 and 300 GeV.

One of the most interesting results in fundamental physics from the 
existing telescopes is a
limit, given by the Whipple collaboration \cite{Whipple}, on the
Quantum Gravity scale to be larger than $4 \times 10^{16}$ GeV 
(about 1/250th of the Planck mass) at the 95 \% confidence level. 
In that study, the data from a TeV gamma-ray flare associated with 
the active galaxy Markarian 421 observed on 15 May 1996 was used to 
place bounds on the possible energy-dependence of the speed of light
in terms of an effective scale for quantum gravitational effects. 

The basic idea behind the measurement is that gamma rays traversing
cosmological distances should notice the quantum fluctuations in
the gravitational vacuum which unavoidably should happen in any
quantum theory of gravitation. These fluctuations may occur on scale 
sizes as small as the Planck length $L_P\simeq
10^{-33}$ cm or time-scales of the order of $t_P \simeq
1/E_P$ ($E_P \simeq 10^{19}$ GeV).

These gammas will therefore experience a ``vacuum polarization'' correction
which should be very small ($O(E/E_{QG})$ where $E$ is the gamma
energy and $E_{QG}$ is
an effective scale for Quantum Gravity, which might be as large as
$E_P$) but might become measurable
after the gamma has traversed cosmological distances. 
In this Quantum Gravity scenario, the requirement of
violation of the Lorenz-Invariance symmetry \cite{Coleman,Nature} emerges 
naturally, providing an
energy-dependent propagation 
speed for electromagnetic waves. Therefore, gammas of different 
energies being emitted simultaneously by a distant source should reach 
our observatories at different times.

In fact, for practical purposes, this delay with respect to the
ordinary case of an energy-independent speed c for massless particles,
for a source at a distance L can be expressed as \cite{Nature}

\begin{equation}
\Delta t \sim \xi \frac{E}{E_{QG}} \,\, \frac{L}{c}
\end{equation}

This delay should be one of the cleanest signatures that Cherenkov 
telescopes are able to study in order to spot Quantum Gravity effects.
Nevertheless, as it has already been pointed out in the literature, it
might be mimicked by effects related to the source physics (internal
production delays for higher energy gammas) or propagation (time
delay due to cascading in intergalactic magnetic fields \cite{Plaga}).
Therefore it will be mandatory to use the scaling of the effect with redshift
to distinguish between Quantum Gravity effects and any source-dependent
phenomena \cite{Whipple}.

The new generation of telescopes, should have an improved sensitivity
and lower threshold. The first characteristic might help in resolving
faster time structures while the second one should allow to observe
sources at much longer distances because of a smaller absorption in the
intergalactic medium. Therefore, some clear improvement 
in this kind of measurements should be expected. 

The goal of this work is trying to estimate how these characteristics 
would actually help in improving the quest for Quantum Gravity
effects. For that, this writeup is organized as follows: first, a very
brief overview of the theoretical framework used in this study is
presented. Second, the results of the expected delay as a function of 
the gamma
ray energy and the source redshift location for different Quantum
Gravity scenarios is discussed. Third, the effect of the existence 
of a Gamma Ray Horizon, including consistently
the effect of the Lorenz Invariance Violation, is shown. 
After that, a discussion about the observational
picture when all these effects are put together is presented and
finally, in the conclusions we give our prospects for these
observations with the coming generation of Cherenkov telescopes.

\section{Overview of the theoretical framework}

When developing any model for Quantum Gravity, it appears naturally the 
necessity to modify some of the most basic continuous symmetries of 
space-time, such as Lorentz Invariance 
\cite{Gonzalez-Mestres,Coleman,Nature,Amelino-Cosmic}. 
In this analysis we shall focus in the violation of the Lorentz Invariance 
symmetry (LIV) caused by Quantum Gravity because it is the responsible 
of the main changes in the kinematics in which is based this study.

The violation of the Lorentz Invariance symmetry (LIV) modifies the 
dispersion relation giving the propagation speed
for gamma rays in a theory-dependent manner. However, since in all
plausible approaches the actual effects are expected to be small, they
can be studied from a phenomenological point of view using an
expansion in terms of the gamma energy divided by the effective
Quantum Gravity scale. Therefore, the actual gamma dispersion relation 
for a massless particle can be expressed in leading order 
as \cite{Amelino-Are}
 
\begin{equation}
\label{dispersion}
E^{2}-c^{2}\vec{p}\,^{2} \simeq E^{2} \xi \left( \frac{E}{E_{QG}} \right)^{\alpha}
\end{equation} 

where $E$ and $\vec{p}$ denotes the energy and the momentum of gammas, 
$\xi$ and $\alpha$ are the LIV free parameters and $E_{QG}$ is the energy
scale for Quantum Gravity.

In this scenario, gamma rays travelling in vacuum can be seen like 
traversing a material medium \cite{Latorre}. Hence, the propagation speed 
of gamma rays should be computed as 
  
\begin{equation}
v = \frac{dE}{dp} = c \left[ 1 + \xi \, \,
\frac{1+\alpha}{2} \left( \frac{E}{E_{QG}} \right)^{\alpha} \right]
\end{equation}

where $E$ is actually the gamma comoving energy. In terms of the
measured gamma energy at the Earth $E_{\gamma}$, the velocity when the
gamma is at a redshift $z$ is

\begin{equation}
v = \frac{dE}{dp} = c \left[ 1 + \xi \, \,
\frac{1+\alpha}{2} \left( \frac{E_{\gamma}(1+z)}{E_{QG}}
\right)^{\alpha} \right]
\end{equation}

Therefore, the actual time-of-flight for a gamma ray coming from a source at 
redshift $z$, is given by

\begin{equation}
t = \int_0^z \frac{c}{v} \frac{dt}{dz} dz 
\end{equation}

where

\begin{equation} 
\frac{dt}{dz}=\frac{1/(1+z)}{H_{0}[\Omega_{M} (1+z)^{3}
+ \Omega_{K} (1+z)^{2} + \Omega_{\Lambda}]^{1/2}}
\end{equation}

being $H_0$ the Hubble constant and $\Omega_M$, $\Omega_K$ and 
$\Omega_{\Lambda}$\footnote{In this analysis we have used the current best
fit values for the fundamental cosmological parameters \mbox{$H_0
= 68 \pm 6 \,\, km \, s^{-1} \,Mpc^{-1}$}, \mbox{$\Omega_M = 0.35 \pm
0.1$} and \mbox{$\Omega_{\Lambda} = 0.65 \pm 0.15$} following
reference \cite{bestfit}.} the standard cosmological parameters.

For practical purposes, we need the difference in the time-of-flight for
gamma rays of different energies ($E_{\gamma}$ and $E'_{\gamma}$) 
produced at the same redshift $z$ 
because this is the signature that Gamma Ray Telescopes are able 
to study. This difference is given by

\begin{eqnarray}
\label{delta-time-of-flight}
\Delta t = t_{E_{\gamma}} -t_{E'_{\gamma}} & = &
\int_0^z \left( \frac{c}{v_{E_{\gamma}}} - \frac{c}{v_{E'_{\gamma}}} \right)
\frac{dt}{dz} dz  \nonumber \\
 & \simeq &  -\xi \, \, \frac{1+\alpha}{2} \, \, 
\frac{E_{\gamma}^{\alpha}-E_{\gamma}^{'\alpha}}{E_{QG}^{\alpha}}
\int_0^z (1+z)^{\alpha} \frac{dt}{dz} dz
\end{eqnarray}

If in the above expressions we take $\alpha=1$,
$E_{\gamma} >> E'_{\gamma}$ and the limit
$z<<1$, we recover the simple expression used by the Whipple collaboration.
 
\begin{equation}
\Delta t \sim \xi \frac{E}{E_{QG}} \,\, \frac{z}{H_{0}}
 = \xi \frac{E}{E_{QG}} \,\, \frac{L}{c} 
\end{equation}

\section{The expected delay}

The free parameters used in our phenomenological approach are
the ones used in references \cite{Amelino-Are}, namely $\xi$, $\alpha$,
and $E_{QG}$. $\xi$ is only a sign ambiguity and hence its possible values 
are $\xi =\pm 1$. Out of the two possible values, in this work we have 
chosen to use $\xi=-1$ because it fits better the present experimental 
constraints \cite{Amelino-Cosmic,Amelino-Planck}.
In what concerns $\alpha$, in Quantum Gravity theories corrections
going like $(E/E_{p})^{\alpha}$, where $E_p$ is the Planck energy, 
typically appear as leading order of more complex
analytic expressions \cite{Amelino-Are,Amelino-Planck}. This motivates us to
study the cases $\alpha = 1$ and $2$.
Nevertheless, $\alpha = 2$ means an $E_{QG}$ suppression factor in the 
difference of time of
flight. This would lead to a non measurable $\Delta t$ ($<10^{-15}
s$). Therefore we cannot exclude the possibility that the leading term of
LIV goes as $(E/E_{QG})^{2}$ but, if it was the case, the Gamma Ray
Telescopes would not have any chance to test it with the gamma
time-of-flights. That forces us to study only the most favourable 
case (potentially largest Quantum Gravity effects) hence $\alpha = 1$.

In this scenario equation~\ref{delta-time-of-flight} is reduced to
 
\begin{equation}
\Delta t \sim \frac{\Delta E}{E_{QG}} 
\int_0^z (1+z) \frac{dt}{dz} dz 
\end{equation}

\begin{figure*}
\begin{center}
\mbox{ \epsfig{file=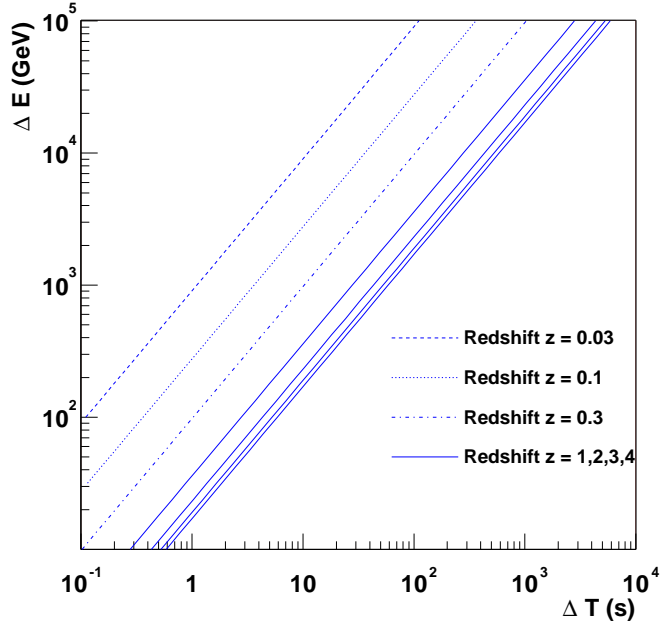,width=9cm} }
\end{center}
\caption[]
{Expected delay as a function of the gamma
ray energy for different redshift sources and a value of \mbox{$E_{QG}$ = $E_p$}.}
\label{EvsT_1_without}
\end{figure*}

Figure~\ref{EvsT_1_without} shows the lines in the plane $\Delta E$ 
versus $\Delta t$
provided by the above equations for a set of different source
redshifts.  From this plot it is clear that, for a given energy
difference between the detected gammas, being capable of observing
sources at larger redshifts should allow to check the $E_{QG}$
 with less demanding time resolutions. Or conversely it
should allow, for a fixed time resolution, to use lower energy gammas to
check the $E_{QG}$. This should help because, given the steep
power-law energy spectrum of all the known high energy gamma
astrophysical sources, lowering the gamma energy scale should allow to
observe a larger number of gammas and, hence, to improve the
experimental accuracy. 

Nevertheless, in spite of the fact that the effective distance
integral giving the $z$ dependence of the time delay has a correction of
$(1+z)$ to the lookback time integrand (which increases the effective gamma
propagation time) it is clear from figure~\ref{EvsT_1_without} that 
this dependence on the redshift saturates quite quickly.

\begin{figure*}
\begin{center}
\mbox{ \epsfig{file=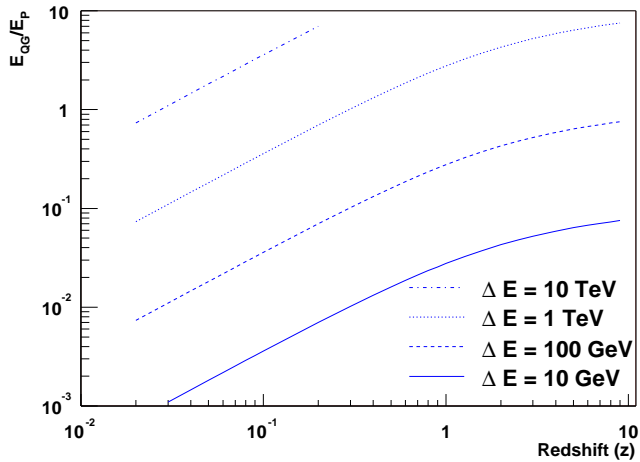,width=9cm} }
\end{center}
\caption[]
{$E_{QG}/E_p$ versus the redshift $z$ of the gamma ray source for
different $\Delta E$ values and fixed $\Delta t$ = 10 s.}
\label{EQGvsZ_T10}
\end{figure*}

Actually the main target of the time-of-flight studies is to explore the
$E_{QG}$ scale. In figure~\ref{EvsT_1_without} it is shown the effect
of observing sources at different redshifts to test a given value of 
$E_{QG}$ ($E_{QG} = E_{p}$). Alternatively, fixing the observable
delay time scale, this result can be shown as in
figure~\ref{EQGvsZ_T10}, as the capability of exploring higher Quantum
Gravity scales as the source redshift increases.

\section{The modified Gamma Ray Horizon}

The above considerations about the benefit of observing higher
redshift sources to explore higher Quantum Gravity effective scales
were implicitly assuming that the gamma rays traverse the
intergalactic space without being altered.
In reality, we expect high energy gamma rays traversing cosmological 
distances to be absorbed by their interaction with 
the diffuse background radiation fields, or ``Extragalactic Background
Light'' (EBL), producing $e^+ e^-$ pairs. The 
$ \gamma_{HE} \gamma_{EBL} \rightarrow e^+ e^- $ cross section is 
strongly picked to $E_{CM} \sim 1.8 \times (2 m_e)$ and therefore, 
there is a specific range in the EBL energy which is ``probed'' 
by each gamma ray energy \cite{Vassiliev}.

This effect should lead to the existence of a ``Gamma Ray Horizon'',
limiting the feasibility of observing very high energy gamma rays
coming from very far distances. The actual value of this horizon 
distance for gamma rays of a given energy, depends on the number 
density of the diffuse background radiation of the relevant energy 
range, which is traversed by the gamma rays. In the range of gamma 
ray energies which can be effectively studied by the next generation 
of Gamma Ray Telescopes (from, say, 10 GeV to 50 TeV), the most 
relevant EBL component is the infrared contribution.

\begin{figure*}
\begin{center}
\mbox{ \epsfig{file=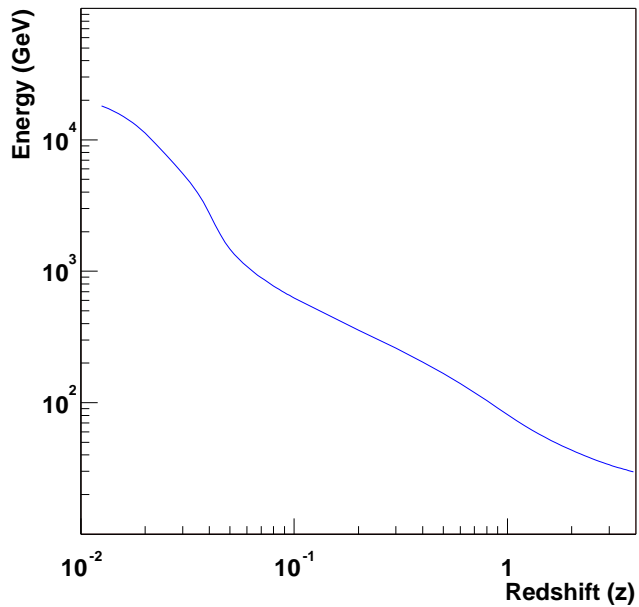,width=9cm} }
\end{center}
\caption[]
{The Gamma Ray Horizon computed assuming purely Standard Model
Physics and using the EBL model presented in the reference 
\cite{Kneiske-Model}.}
\label{Hor_classic}
\end{figure*}

Quite precise predictions \footnote{In spite of the precision of 
these predictions they have 
still rather considerable systematic uncertainties due to the 
poor knowledge of some of their ingredients (direct measurement of the
EBL at small redshifts and theoretical extrapolation to large redshifts) 
and are therefore quite inaccurate. This situation is, nevertheless, improving
quite quickly with a whole harvest of new measurements.} of the Gamma 
Ray Horizon have
been made but, unfortunately, so far no clear confirmation can be 
drawn from the observations of the present generation of Gamma Ray 
Telescopes. In this work we have used the approach and procedures
described in \cite{Vassiliev,Horizon,Kneiske-StarEvo} which predict 
a Gamma Ray Horizon as shown in figure~\ref{Hor_classic}.

Nevertheless, the above prediction has been obtained by assuming purely
Standard Model Physics. Recent claims of observation of very high
energy gamma ray events coming from Mkn 501 \cite{Mkn501TeV}, a blazar
at redshift $z \sim 0.03$ which would somehow contradict the above 
prediction of the Gamma Ray Horizon, have motivated some authors 
\cite{Amelino-Planck,Sato,Aloisio,Kifune} to revise that prediction. 
It has been pointed out that
the actual existence of LIV should also affect the calculation
of the Gamma Ray Horizon, since the threshold energy of the EBL needed
to produce $e^+ e^-$ pairs is modified \cite{Amelino-Planck}. This
effect could explain the very high energy observed events.

Therefore, to be consistent with the framework of our analysis, we've also 
undertaken the task of implementing consistently in the complete
calculation of the Gamma Ray Horizon the inclusion of the
threshold-modifying effects coming from the effective Quantum Gravity
inspired dispersion relations discussed in the previous sections. For
that we've followed the approach of reference \cite{Amelino-Planck},
where the modified threshold value for the gamma ray momentum to
produce $e^+ e^-$ pairs is deduced\footnote{In \cite{Amelino-Planck},
the notation is slightly different. There appears a parameter $\eta$,
which is \mbox{$\xi \cdot (E_{p}/E_{QG})^{\alpha}$}}

\begin{equation}
\label{p_threshold}
p_{th} = \frac{m_e^{2}}{\epsilon} 
+ \xi \, \, \frac{p_{th}^{2+\alpha}}{4 \epsilon
E_{QG}^{\alpha}} \left( \frac{1}{2^{\alpha}}-1 \right)
\end{equation}
  
where  $\epsilon$ is the energy of the EBL gammas
and $p_{th}$ it the threshold momentum for the gamma ray. 

In the standard calculation of the Gamma Ray Horizon \cite{Stecker-95},
the optical depth for gamma rays of energy $E_{\gamma}$ coming from a
source at redshift $z$ is obtained from the integration

\begin{equation}
\tau(E_{\gamma},z) =
\int_{0}^{z}dz'\frac{dl}{dz'}\int_{0}^{2}dx\frac{x}{2}\int_{\epsilon_{tr}}^{\infty}d\epsilon\cdot
n(\epsilon,z') \sigma[2xE_{\gamma}\epsilon(1+z')^{2}]
\end{equation}

where \(x \equiv 1-cos\theta\) being $\theta$ the gamma-gamma
scattering angle,
$n(\epsilon,z')$ the spectral density of the EBL gammas $n(\epsilon)$
at the given z' and $\epsilon_{th}$ the threshold energy for the EBL
gammas which is given by

\begin{equation}
 \epsilon_{tr} = \frac{2m_e^{2}}{E_{\gamma} x(1+z')^{2}}
\end{equation}

After we include LIV, we arrive to a modified threshold condition of

\begin{equation}
 \epsilon_{tr} = \frac{2m_e^{2}}{E_{\gamma} x(1+z')^{2}} 
+ \xi \, \,  \frac{2}{x(1+z')} \, \, \frac{[E_{\gamma}(1+z')]^{1+\alpha}}{4
E_{QG}^{\alpha}} \left( \frac{1}{2^{\alpha}}-1 \right)
\end{equation}

where $E_{\gamma}$ and $\epsilon_{tr}$ are the energies at the Earth
and, therefore, we had to add some $(1+z')$ factors respect to the
equation~\ref{p_threshold}. The dependence on the scattering angle has
also been introduced as a global factor $2/x$.

\begin{figure*}
\begin{center}
\mbox{ \epsfig{file=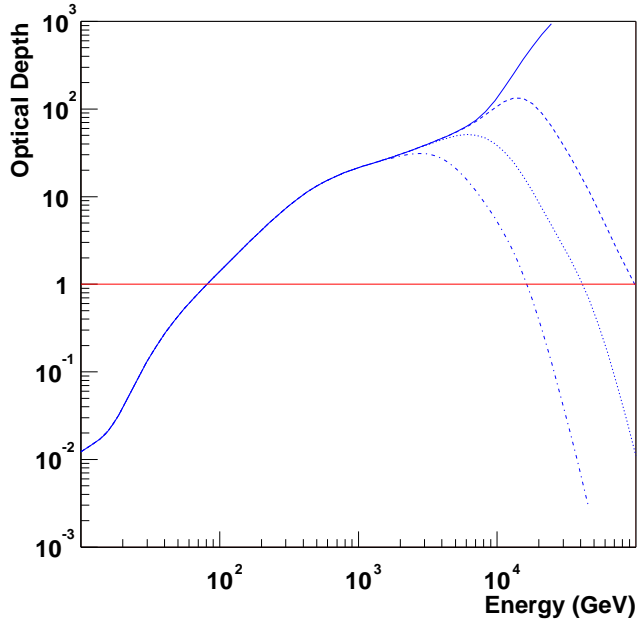,width=9cm} }
\end{center}
\caption[]
{The Optical Depth as a function of the gamma ray energy for a 
redshift \mbox{$z$
= 1}.The continuous line is
for a calculation without LIV whereas the dashed, dotted and
dash-dotted lines are
for values of $E_{QG}$ = $E_p$, $E_p$/10 and $E_p$/100 respectively. 
The horizontal line shows that the Gamma Ray 
Horizon is defined for the energies were the Optical Depth is equal to one.}
\label{OptDepth}
\end{figure*}

The results comparing the optical depths with and without this
threshold correction (once again in the $\alpha =1$ and $\xi = -1$ 
scenario) are shown in figure~\ref{OptDepth}. For these parameter
values the net effect of the correction is increasing the threshold
energy in such a way that, for any given gamma energy, harder EBL
photons are responsible for their absorption. 

In the Optical Depth calculation without LIV, the gamma rays of 
$10^5$ GeV ``probe'' the EBL spectrum at around  $\lambda_{EBL} 
\sim 100 \mu m$ where the density has a maximum, while when the LIV
correction is present, they probe the EBL spectrum at lower
wavelengths, where the value of the density is smaller
\cite{Horizon}. The main consequence of this is that 
in the calculation with LIV corrections the universe becomes transparent
again for large energy gammas whereas the results for the lower energy 
gammas of the plots remain basically unchanged. The correction for large energy
gammas is so severe that avoids the existence of a Gamma Ray Horizon
for moderate redshift sources, like Mkn 421, giving then a plausible
explanation for the observed gamma events as has already being 
mentioned.

\begin{figure*}
\begin{center}
\mbox{ \epsfig{file=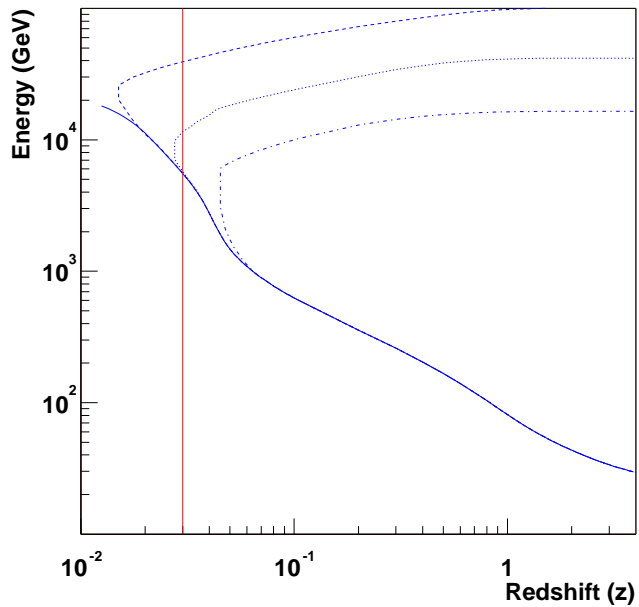,width=9cm} }
\end{center}
\caption[]
{The Gamma Ray Horizon computed assuming LIV. The continuous line is
for a calculation without LIV whereas the dashed, dotted and
dash-dotted lines are
for values of $E_{QG}$ = $E_p$, $E_p$/10 and $E_p$/100 respectively.
Notice that for  $E_{QG}$ $\sim$ $E_p$/10 the Gamma Ray Horizon disappears 
for the redshift of the observed blazars Markarian 421 and 501 ($z\sim0.03$), 
which is shown in the plot with a vertical line.} 
\label{Hor_LID}
\end{figure*}

The obtained Gamma Ray Horizon is shown in figure~\ref{Hor_LID} for different
$E_{QG}$ and compared with the ``standard'' one. If one
assumes high $E_{QG}$, it is clear from that plot that the main
difference due to LIV should happen for low
and moderate redshift sources whereas for large redshift sources no
effect should be observable in the Gamma Ray Horizon. On the other
hand for not so high $E_{QG}$, the universe becomes transparent again
for a gamma ray energy that can be reached with current AGN models 
\cite{Mannheim} (15 TeV for $E_{QG} = E_{p}/100$). The fact that
$\gamma$ rays of this high energies have been detected for close
objects ($z = 0.03$) \cite{Mkn501TeV} but not for
objects at larger redshift suggests that the $E_{QG}$ must be in excess 
of $\approx E_{p}/100$. Actually, one of the most recent results about 
the effects of LIV in the Gamma Ray Horizon was presented in 
\cite{Stecker-Glasow} where TeV gamma ray observations of the active 
galaxy Markarian 501 were used to set a lower limit on the Lorentz 
Invariance breaking parameters.

\section{Discussion}

We shall now put together the effects discussed in the previous sections 
to get a complete picture of the expectations. As already
mentioned, on the one hand the capability of observing more distant 
sources should allow to have more lever arm to explore higher
effective Quantum Gravity scales. Unfortunately, on the other hand,
for more distant sources, the gamma absorption in the EBL is stronger
and the Gamma Ray Horizon happens at smaller gamma energies. 

\begin{figure*}
\begin{center}
\mbox{ \epsfig{file=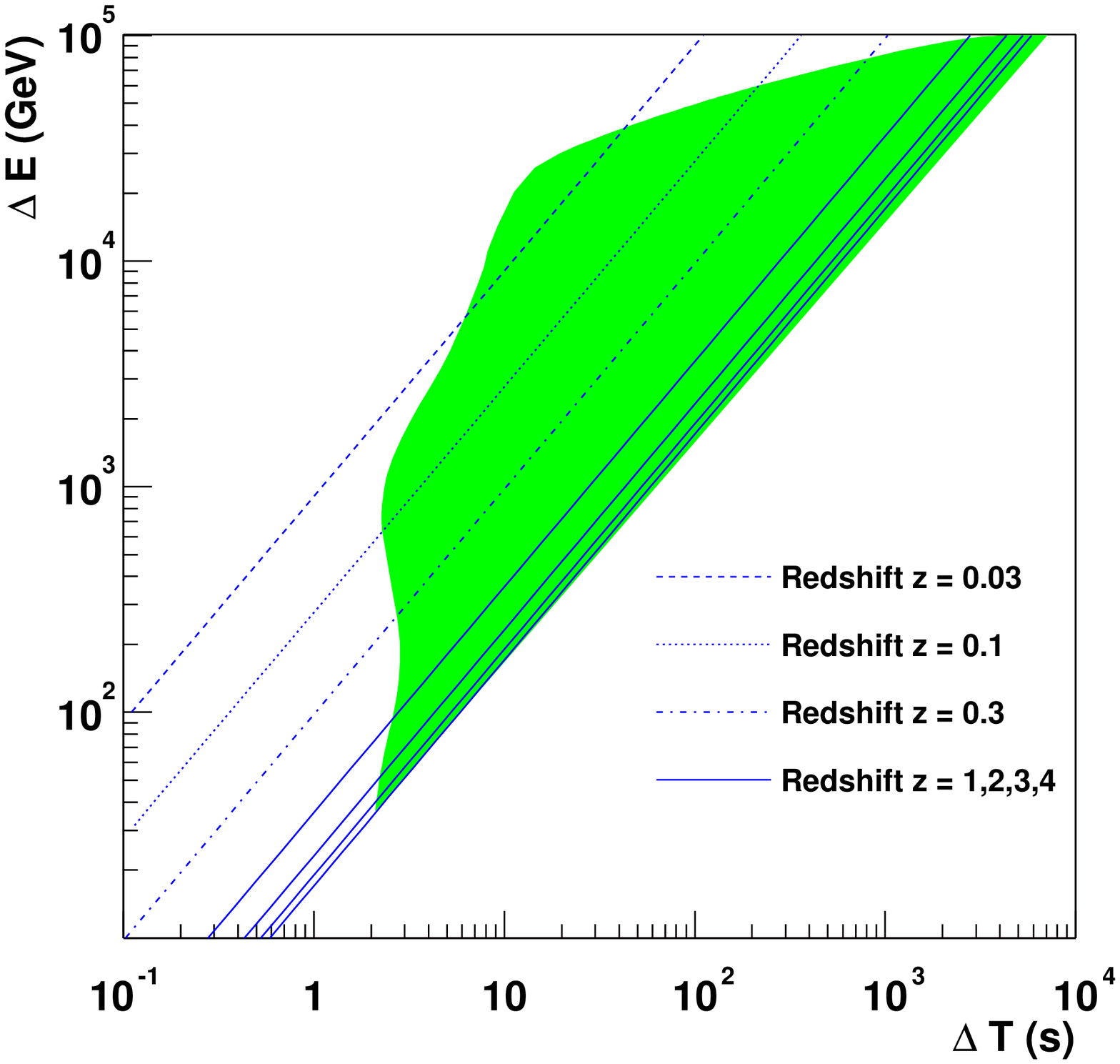,width=6.5cm} }
\mbox{ \epsfig{file=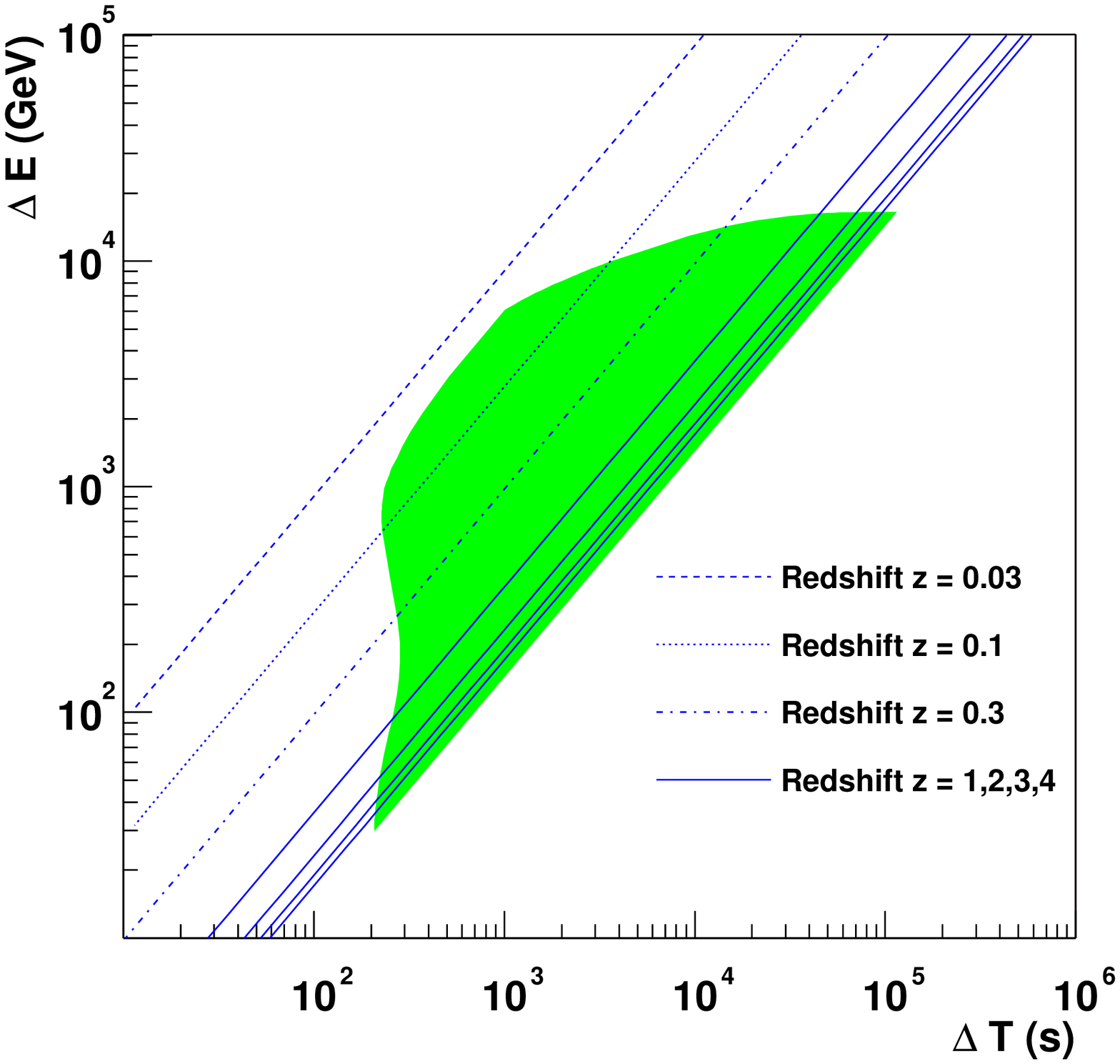,width=6.5cm} }
\end{center}
\caption[]
{Expected delay as a function of the gamma ray energy for different
redshift sources. The shadow area is a projection of the Gamma Ray
Horizon in the $\Delta E$ v.s. $\Delta t$ plane. Therefore, the
possible Quantum Gravity effects can hardly be studied in this area due to
the lack of gamma rays. $E_{QG}$ = $E_p$ and $E_p$/100 for the left and
right plots respectively. Notice the change 
in the time scale axis for the different plots.}
\label{EvsT}
\end{figure*}

This trade-off is clearly summarized in figures~\ref{EvsT} where, on 
top of the 
$\Delta E$ versus $\Delta t$ lines predicting the propagation delays, one
can see the parameter region beyond the Gamma Ray Horizon. It must be
stressed here that the Gamma Ray Horizon does not mean a ``hard
boundary'' since it just gives the condition for an e-fold reduction
of the observed flux. Because of that, the shaded area given by the
Gamma Ray Horizon has to be understood as the region in which the flux
reduction due to the absorption starts being strong enough to make the
source observation difficult.


From these figures it is clear that the time resolution needed depends 
strongly on the actual effective Quantum Gravity scale that one would like 
to test. To explore up to the Planck mass, 
time resolutions on the scale of just few seconds will be mandatory if 
gamma rays of up to few hundred GeV are used (or few tens of seconds 
if gammas of few TeVs are used) quite independently on the actual source 
redshift. But resolution of several minutes could be enough for a scale of 
one hundredth of the Planck mass.

In view of these figures it is not obvious how much having a reduced
threshold would help the next Cherenkov Telescope generation to
improve in this kind of analysis. In fact, from the above figures the
straight conclusion is that the observation of the nearby 
Blazars (Mkn 421, at $z=0.031$ is actually the closest known Blazar) 
in an energy range of $10^3-10^4$ GeV seems to be the most comfortable
scenario for exploring the highest effective Quantum Gravity scale.
 
\section{Conclusions}

 In this work the potential of the next generation of Gamma Ray
 Telescopes in improving the existing tests of an effective Quantum
 Gravity scale from the study of the propagation delay for gamma rays
 of different energies coming from a distant astrophysical source has
 been analyzed. These new telescopes should have better sensitivity
 and lower threshold than the present ones, enabling a better time 
 resolution and the observation of sources at higher redshift and
 therefore, a clear improvement in this kind of measurements should
 be expected.

 Detailed calculations on the actual expected delay for sources at any
 redshift have been presented. They show that the effective delay time grows
 with the redshift faster than the lookback time. In spite of that,
 the redshift dependence of the relationship between the gamma energy 
 scale and the observable arrival time delay has been shown to saturate
 rather quickly as the source redshift increases.

 In addition, a calculation of the Gamma Ray Horizon due to the gamma
 absorption in the extragalactic background light (EBL) taking
 consistently into account the Lorenz Invariance Violation (LIV)
 effects has been carried out. This absorption should limit the
 observability of high energy gammas coming from very distant
 sources but, as it has already been realized in the past, the 
 inclusion of LIV effects opens a window for very high energy gammas.

 Putting all these effects together, the global picture is
 that quite stringent requirements on the time resolution (in the
 scale of few seconds) should be
 achieved to be able to explore an effective Quantum Gravity scale
 close to the Planck mass. In fact, given the effect of the Gamma Ray
 Horizon, this demanding time resolution constraint cannot be avoided
 by using large redshift sources unless very high energy gamma rays are
 used. Although, this resolution is less restrictive
(several minutes) for a scale around one hundredth of the Planck mass.
In view of this situation, maybe the observation of the nearby 
Blazars in an energy range of $10^3-10^4$ GeV could be the most comfortable
scenario for exploring the highest effective Quantum Gravity scale.
Given this result, the new generation of telescope arrays might be the 
most promising choice since they can provide a much larger collection 
area for higher energy gamma rays.

\begin{ack}

We would like to thank our colleagues of the MAGIC collaboration for
their comments and support. We are indebted to S.M.Bradbury from
the VERITAS collaboration for triggering our interest in this
subject. We thank R.Pascual for discussions and comments on this manuscript. 
We'd like also to thank J.Cortina, O. de Jagger, 
J.Garriga, A.Grifols, E.Lorenz, K.Mannheim and A.Saggion 
for their encouraging comments. And finally, we'd like to thank to 
O. Bertolami, G. Amelino-Camelia and F. Stecker for their comments 
in the preprint stage of this article.

\end{ack}

\end{document}